# Discussion on Nonlinear Dynamic Behavior of Suspension Based Bridge Model

Marcus Varanis[a], José Manoel Balthazar[c], Felipe Lima de Abreu[b], Pedro Augusto Beck[b], Mauricio Aparecido Ribeiro[c], Clivaldo de Oliveira[b]

*marcus.varanis@ufms.br*

[a]*Physics Institute, Federal University of Mato Grosso do Sul*
[b]*Faculty of Engineering, Federal University of Grande Dourados*
[c]*Federal Technological University of Paraná*

## Abstract

In this paper we explore the numerical study. of the Nonlinear Behavior of Suspension Bridge Models. The study of suspension bridges is one of the classic problems of mechanical vibrations, one of the most famous collapses of which was that of the Tacoma Narrows Bridge. This paper covers an initial explanation of vibrations in a suspension bridge. To do this, three different systems are going to be simulated: The first being a asystem where only the vertical vibrations of the bridge deck are taken into account, the second covering the vibrations of the main cable and the roadbed, and lastly, a system that takes both vertical and torsional vibrations into account. A time-frequency analysis will also be done on all systems with temporal response, Fast Fourier Transform (FFT) and Continuous Wavelet Transform (CWT), plus in a specific case the use of Hilbert-Huang transform (HHT). Poincaré maps and Lyapunov exponents are used to characterize the dynamics of the system. In particular, in the vertical and torsional system, an explanation of why the Tacoma Bridge oscillations have undergone an abrupt change from vertical to torsional oscillations. Thus, extremely rich dynamic behaviors are studied by numerical simulation in the time and frequency domains.

## 1. Introduction

The study of mechanical vibrations is one of the classic fields of mechanical engineering. Problems of how vibrations impact a material, such as changing the entire vibrational behavior of a simple spring-mass system due to an external force, are grounds for further study. A simple change in initial conditions can lead to abrupt changes in vibration amplitudes, which is one of the characteristics of a nonlinear system, such as pointed by researches made by "Nayfeh" in [1, 2]. Time-frequency analysis (TFA) methods [3] are important to understand the behavior of a mechanical system and how a simple external force, such as wind, can cause a bridge to collapse [4].

One of the cases studied about the oscillations of a mechanical system was the Tacoma Bridge [5, 6], an example that serves even today as a good introduction to the subject. In this case, it was observed that the oscillations of the Tacoma bridge showed certain patterns that did not agree with the linear theory [7]. Lazer-McKenna were among those who presented new models to be able to explain with nonlinear systems the rich dynamics of the bridge [8]. As certain atypical effects were observed in the oscillations of the bridge, such as the change from large amplitudes to small amplitudes with practically the same external force, in addition to the Galloping effect of the bridge cables [9] and an almost abrupt change from vertical to torsional oscillations. Lazer-McKenna devised 3 systems to explain these nonlinearities. The first model was proposed to explain how the center spam oscillations of an idealized suspension bridge occur [10] , which could explain how the bridge presented different forms of oscillations. In the second model [11] the oscillations of the main cables and the roadbed were admitted, which was idealized to describe the movements of both the roadbed and the main cables, a model that could show phenomena such as cable bumping and Galloping cables. Finally, the third model [12] was designed as a cross section of the bridge deck, which would clarify why there was an almost instantaneous change from vertical to torsional oscillations observed on the Tacoma Bridge.

In [13] a new nonlinear model for a suspension bridge is presented with the parameters corresponding to the collapsed Tacoma Narrows Bridge. A long and in-depth study on various mathematical models of suspension bridges can be found in [14, 15]. Several applications on aeroelasticity and the flutter phenomenon in suspension bridges have been proposed [16, 17, 18, 19]. Many applications of structural condition in suspension bridges [20, 21, 22] are based on numerical models.

Numerical simulations are necessary to understand and visualize the phenomena, and to investigate how the conditions of the systems (vertical position, torsional or amplitude of the external force) interfere in the responses of the proposed models. Thus, several situations will be addressed and compared in order to obtain an analysis of the initial conditions of the proposed models, demonstrating the effectiveness of these models that present a rich dynamic in the analysis of the oscillations of a suspension bridge.

In this paper the goal is to perform a review of models already presented in the literature from a numerical roadmap based on fourth order Runge-Kutta scheme (detailed in Figure 1), with main focus on the response in the frequency domain through the Time frequency analysis (TFA). The TFA used in this paper are the wavelet and Hilbert-Huang transform, emerging techniques suitable for the analysis of nonlinear dynamical systems capable of characterizing the chaotic dynamics of the system [23], with the addition of a method of characterization in the frequency domain through the Fast Fourier Transform, suitable for stationary analysis.

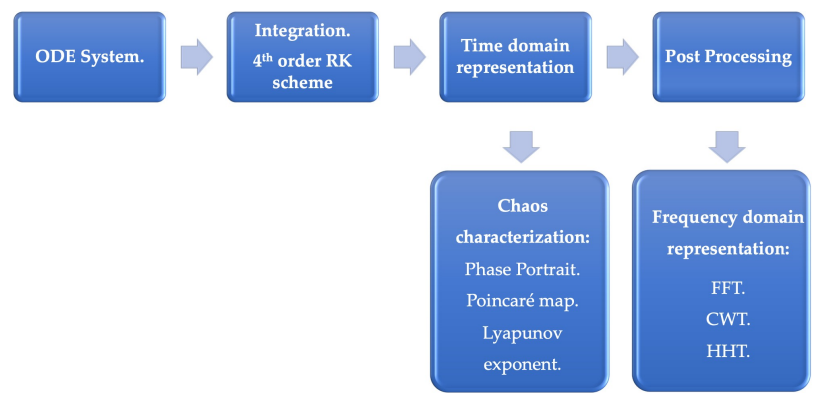

Figure 1. Numerical roadmap.

After the introduction, in section 2, the development of the equations of motion for each model is presented, exploring their particularities. In section 3, the results of each model are presented as follows: Time domain response, Phase portrait, TFA (CWT, HHT), FFT, Poincaré map and Lyapunov exponent. Finally, in section 4, the final remarks of this paper are discussed. This study is a sequence of the results presented in [24].

## 2. Mathematical Background

### 2.1. Lazer-Mckenna suspension bridge model

This model represents the center span oscillations of a suspension bridge from a tower of the bridge to a distance x [10]. The roadbed is treated as a one-dimensional beam with constant stiffness $k$ that answers Hook's Law, and the cable remains a nonlinear spring of stiffness $R$. This model treats the motion only from a point at a distance x from the reference towers. The model was first presented in [8] and detailed [10, 11, 25]. Afterwards, the authors did a brief analysis in the frequency domain and re-presented the model in the paper [24]. As demonstrated in [26, 24], the oscillations of this system can be analogized as a mass-spring-damper system with the addition of an elastic (nonlinear spring) that acts only against extension and has a response that respects Hooke's law of proportionality. The system is given in Figure 2.

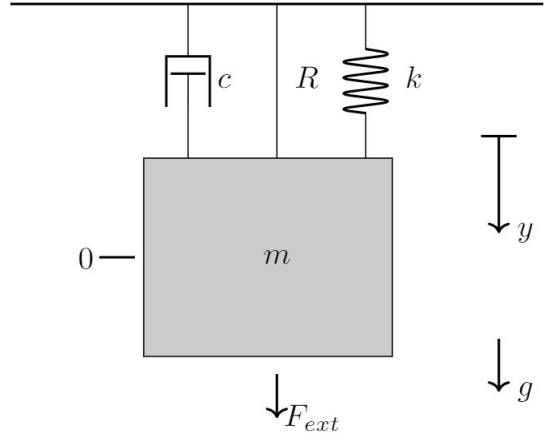

Figure 2. Mass-spring-rubber model, where $c$ is damping coefficient, $k$ and $R$ are the spring and the rubber band constant, $g$ is gravity constant and $F_{ext} = \lambda sin(\mu t)$ is external force.

The equation of motion of this system is presented in Eq.(1).

$$y'' + \delta y' + ay^+ - by^- = g + \lambda \sin(\mu t) \tag{1}$$

Two important points are used in this formulation: the first is the positive y-axis asymmetry being the downward oscillations ($y^+$ as $y > 0$, and $y^-$ as $y < 0$), and the second is the use of the parameter $a$, found at positive displacements ($y^+$), being the sum of the linear spring constant and the nonlinear spring, and the parameter $b$ being only the linear spring at negative displacements ($y^-$). The addition of the elastic element $R$ makes the system exhibit strongly nonlinear response and rich dynamics.

## 2.2. Coupling the motions of the roadbed and the cable

The second model used in this study deals with both the motion of the roadbed and the main cable. Here, the roadbed is modeled as a vibrating beam, but the main cable oscillates relative to its own reference, like a vibrating string. The motion of the roadbed and main cable are coupled by the cable stays and are treated as nonlinear springs. The scheme of the model is represented as shown in Figure 3.

The equations are obtained assuming the following considerations: the coupling is given by a nonlinear spring of stiffness $k$, similar to the first model, where this act only against tension, not exerting force against compression, the roadbed is subject to its own weight and an external force is applied to the main cable due to wind or the movement of the towers. The variable $v$ measures the motion from the equilibrium of the main cable and $u$ measures the motion of the roadbed [11]. The equations of motion and the boundary conditions for the system are given respectively by Eq.(2), (3) and (4).

$$m_1 v_{tt} - T v_{xx} + \delta_1' v_t - k(u - v)^+ = \varepsilon' f(x, t) \tag{2}$$

$$m_2 u_{tt} - EI u_{xxxx} + \delta_2' u_t + k(u - v)^+ = W_0(x) \tag{3}$$

$$u(0) = u(L, t) = u_{xx}(0) = u_{xx}(L, t) = v(0) = v(L, t) = 0 \tag{4}$$

Normalizing Eq.(2) and (3) by the masses ($m_1$ and $m_2$), and taking the approximation by the eingenfunction expansion of the constant function $W_0(x) = W_0 sin(\pi x/L)$, in addition to looking for non-nodal solutions of the form $u(x, t) = y(t)sin(\pi x/L)$, $v(x, t) = z(t)sin(\pi x/L)$ and an external force such as $f(x, t) = g(t)sin(\pi x/L)$, by substituting the considerations into Eq.(2) and (3), the term $sin(\frac{\pi x}{L})$ could be removed from the equation dividing and getting Eq.(5) and (6):

$$z'' + \delta_1 z' + a_1 z - k_1(y - z)^+ = \varepsilon g(t) \qquad (5)$$

$$y'' + \delta_2 y' + a_2 y + k_2(y - z)^+ = W \qquad (6)$$

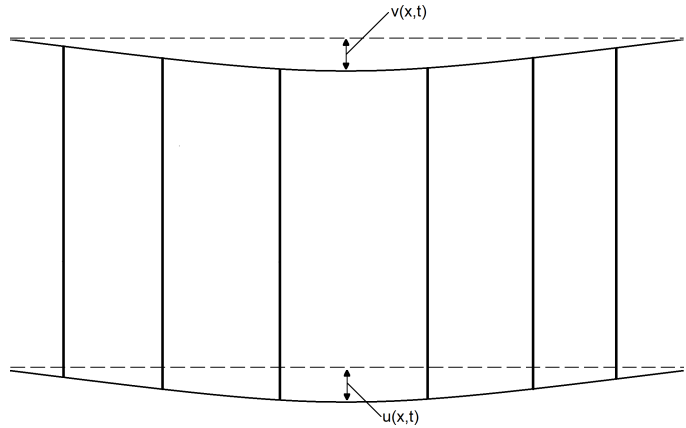

Figure 3. Coupling the motions of the roadbed and the cable, where v(x,t) is the main cable displacement and u(x,t) is the roadbed displacement, both in the vertical direction.

Where, $\delta_1 = \delta_1'/m_1$, $a_1 = (\pi/L)^2 T/m_1$, $k_1 = k/m_1$, $\varepsilon = \varepsilon'/m_1$, $\delta_2 = \delta_2'/m_2$, $a_2 = EI(\pi/L)^4/m_2$, $k_2 = k/m_2$, $W = W_0/m_2$.

Due to the simplifications made in Eq.(5) and (6), the model describes the oscillation of only two points at a distance x from the center span of the bridge, which are similar to the model presented in section (2.1), one point is described on the main cable and the other on the roadbed, being parallel to each other, as proposed in [11].

### 2.3. McKenna torsional oscillation

This model deals with the torsional motion in a suspension bridge, a model that will be used to show one of the nonlinearities found in the Tacoma Narrows bridge: the almost instantaneous transition from vertical to torsional motion [7]. The model is shown in Figure 4.

The derivation takes into account two types of motion (vertical and torsional), which can be obtained by assuming the cross section of the roadbed as a rod suspended on both sides by nonlinear springs with stiffness k, similar to the previous models. The variable $y$ represents the distance down from the center of gravity of the rod without any loading, the angle $\theta$ that the variable $y$ makes with the horizontal, and the extension is given by $(y - l \sin \theta)^+$ in one spring and $(y + l \sin \theta)^+$ in the other, and finally using Lagrange energy methods the following system of equations presented in Eq.(8) is obtained.

$$(1/3)ml^2\ddot{\theta} = (kl)\cos(\theta)[(y - l\sin\theta)^+) - (y + l\sin\theta)^+] \qquad (7)$$

$$m\ddot{y} = -k((y - l\sin\theta)^+) + (y + l\sin\theta)^+) + mg \qquad (8)$$

By simplifying and adding a small viscous damping term in both dimensions ($\delta\dot{\theta}$ and $\delta\dot{y}$), and adding a forcing term in the torsional dimension ($f(t) = \lambda\sin(\mu t)$), have the following system of equations of motion, as shown in Eq. (9) and (10).

$$\ddot{\theta} = -\delta\dot{\theta} + (3k/ml)\cos(\theta)[(y - l\sin\theta)^+) - (y + l\sin\theta)^+] + \lambda\sin(\mu t) \qquad (9)$$

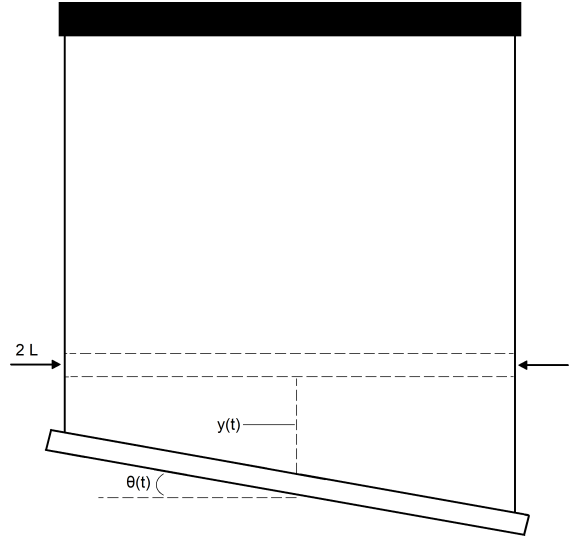

Figure 4. McKenna torsional oscillation model, where y(t) is the vertical displacement, θ(t) the angular displacement and L half size of the bridge deck crossection.

$$\ddot{y} = -\delta\dot{y} - (k/m)[(y - l\sin\theta)^{+}) + (y + l\sin\theta)^{+}] + g \qquad (10)$$

A further formulation of the development of Eq.(8), (9) and (10) is given in [12].

## 3. Results and discussions

The model analyses follow the same methodology. First, the simulation time was 3000 seconds, but with the intention of analyzing without the vibration transient period, the time is within the interval ($2500 \geq t \geq 3000$) seconds. For solving the equations of motion, the fourth order Runge Kutta method was used in Python language with the help of its scientific libraries, NumPy, SciPy and Matplotlib [27, 28]. In all cases also, the time response and phase space will be used for analysis in the time domain. Fast Fourier Transform (FFT), Continuous Wavelet Transform (CWT) and Hilbert-Huang Transform are used, for analysis in the frequency domain [3, 23].

### 3.1. Case 1 - Lazer-Mckenna suspension bridge model

The first model (Eq.(1)) aims at explaining how a suspension bridge can oscillate considering how a beam with relatively low stiffness and with cable stays that correspond to nonlinear springs that have a higher order stiffness can make the dynamics of the bridge respond in rich dynamics from different amplitudes of external forces. Two Analyses will take place: the Analyse for ($\lambda = 7$ N/kg) and ($\lambda = 15.5$ N/kg). As nomenclature for the parameters viscous damping is used as $\delta$, linear spring constant and nonlinear spring constant as $a$, linear spring constant $b$, all normalized by the mass $m$, $g$ denotes the acceleration of gravity and $\mu$ the frequency of the external force. The values for the parameters used will be as follows: $\delta = 0.01$ [Ns/mkg], $a = 13$ [N/mkg], $b = 2$ [N/mkg], $g = 9.81$ [m/(s²)] e $\mu = 0.77$ [rad/s], the selected parameters were based on references [24, 26].

We calculate the Lyapunov Exponents with the sweep of the parameter $\lambda \in [6, 16]$, together with the bifurcation diagram of the displacement $x$. Figure 5(a) represents the Lyapunov exponents for $\lambda \in [6.16]$ considering $\delta = 0.01$ [Ns/mkg], $a = 13$ [ N/mkg], $b = 2$ [N/mkg], $g = 9.81$ [m/(s²)] and $\mu = 0.77$ [rad/s] and Figure 5 (b) represents the bifurcation diagram for displacement. We can see that for the intervals of $\lambda$ in [15.44, 15.46], [15.49, 15.52], [15.54, 15.56], [15.66, 15.71] and [15.73, 16] the behavior of Eqs. (1) has a chaotic behavior, however, for the other values of the interval the system behaved periodic.

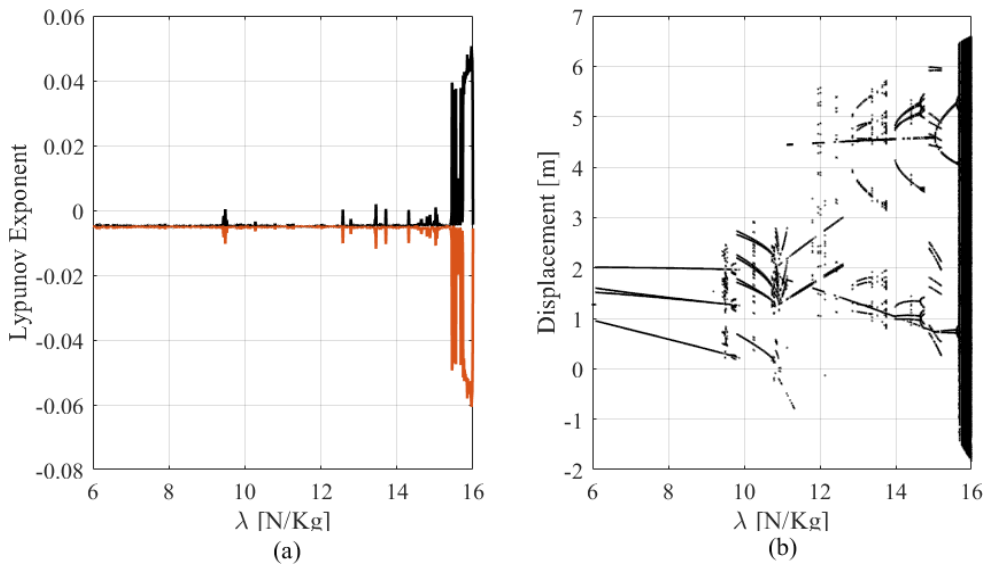

Figure 5. (a) Lyapunov Exponent to $\lambda \in [6, 16]$ and (b) Diagram Bifurcation to displacement.

We calculate the Lyapunov exponent (MLE), the figure 6 (a) represents the parameter space of $a \in [6, 1.6]$ and $b \in [1, 2]$. The parameters $a$ and $b$ are related to the positive displacement ($y^+$) (sum of positive displacement) and ($y^-$) (negative displacement), respectively. And Figure 6 (b) represents the space of $\lambda \in [6, 16]$ $\mu \in [0.7, 1.0]$, that is parameter of the external force applied in the system, i.e., $\lambda$ is amplitude and $\mu$ is frequency of the external force. In both cases we can observe that the region in black (negative MLE) represents the periodic behavior of eq.1 and the region from yellow to red the chaotic behavior (negative MLE).

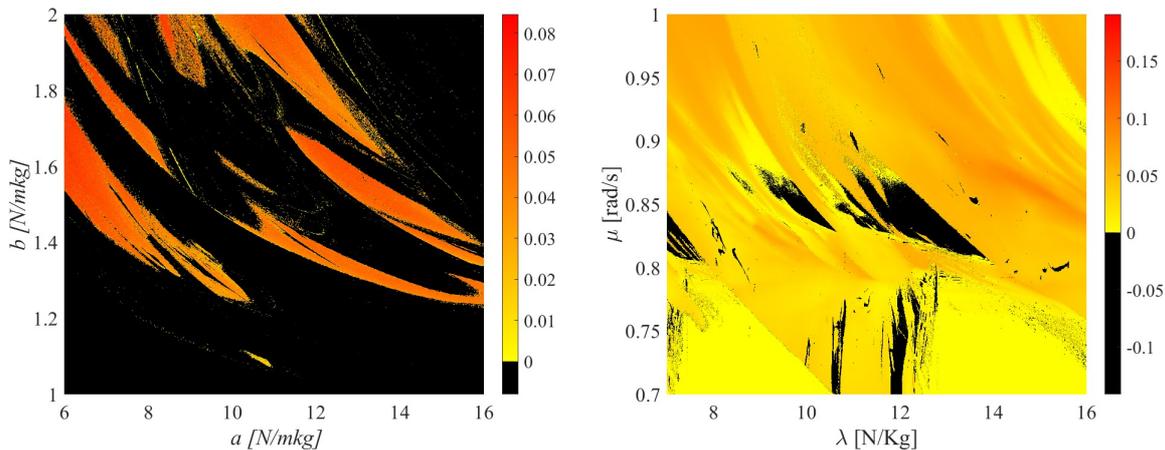

Figure 6. Maximum Lyapunov Exponent. (a) $a \in [6, 16] \times b \in [1, 2]$ and (b) to $\lambda \in [6, 16] \times \mu \in [0.7, 1]$

In addition to the time-frequency analysis tools, the FFT and CWT, the Hilbert Huang Transform (HHT) and the Poincaré Map will be used [29].

Thus, the results of the time domain analysis are presented in Figure 7, while the results of the frequency domain analysis are presented in the Figures 8 e 9.

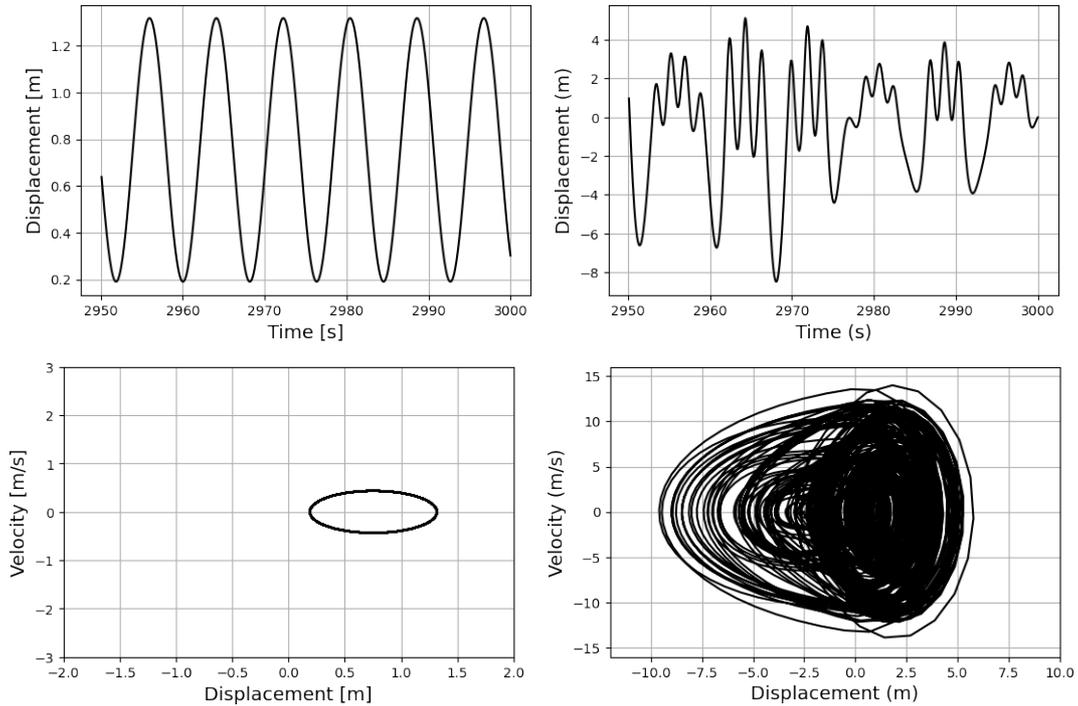

Figure 7. (a)Time domain response(λ=7), (b)Time domain response (λ=15.5), (c)Phase Portrait (λ=7), (d)Phase Portrait (λ=15.5).

In the first analyse, where the parameter of the external force amplitude has a value of 7 [*N/kg*], Figures 7 ((a),(c)), Figures 8 ((a),(c)) and Figure 9 (a) show that the system presents a linear, stationary response. For the second analyse, a significant increase in the amplitude of the external force, to 15.5 [*N/kg*], caused a completely different response, as presented in the Figures 7 ((b), (d)), besides that in Figure 8 (b), a rich dynamic is contemplated and presents the appearance of harmonics, phenomenon known as frequency scattering visualized in the analyses through CWT (Figure 8 (d)) and HHT ((Figure 9 (b)). According to the changes seen in these results, the amplitude of the external force can interfere greatly in how the system will behave, varying from small to large amplitudes. This was an effect that occurred in the Tacoma Norrows bridge, as a way to solve this problem, a reinforcement was added to the main deck so that the stiffness of the bridge increased, and this problem would not recur for small changes in amplitudes [7].

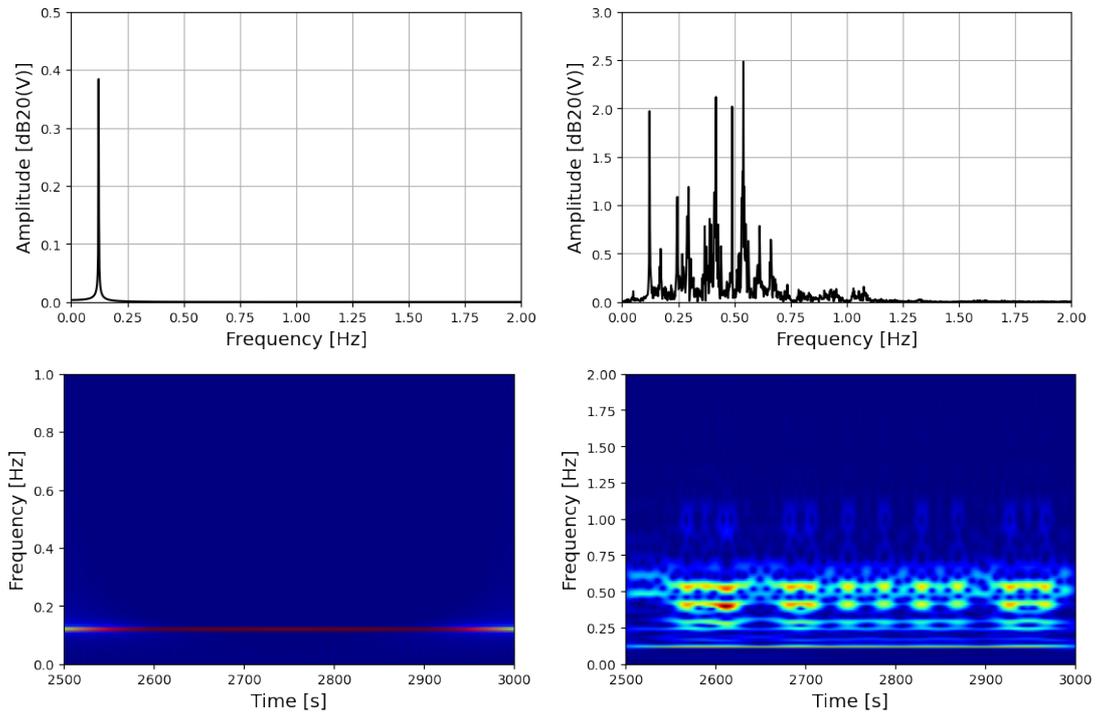

Figure 8. Frequency domain response (a)FFT ($\lambda$=7) (b)FFT ($\lambda$=15.5), (c)CWT ($\lambda$=7), (d)CWT ($\lambda$=15.5)

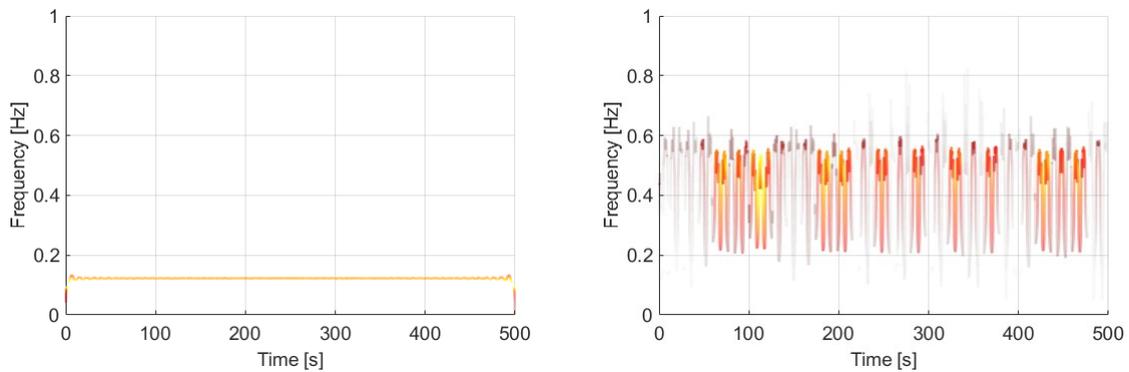

Figure 9. Frequency domain response (a)HHT ($\lambda$=7), (b)HHT ($\lambda$=15.5)

The frequency spectra presented in the Figures 8 (d) and 9 (b) show that the system responds in a nonlinear way and represents a strong indicator of the presence of chaos in the system.

Figure 10 presents the Poincaré Map analysis which further confirms that for the first analyse, the system vibrates at a single frequency (Figure 10 (a)) and for the second analyse, several frequencies are found in the response (Figure 10 (b)).

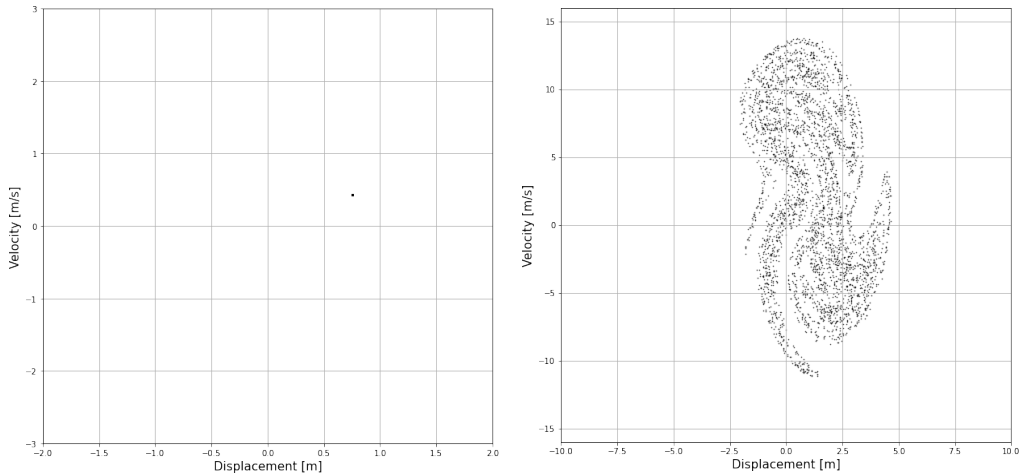

Figure 10. Poincaré map analysis (a)Poincaré map($\lambda$=7), (b)Poincaré map ($\lambda$=15.5).

The numerical results reveal very rich and complex nonlinear behavior of the system.

## 3.2. Case 2 - Coupling the motions of the roadbed and the cable.

In the system shown in Eq.(5) and (6), follows a similar methodology as the previous one, but also consider the motion of the main cable. Thus, from this coupling it is also possible to observe how the dynamic response of the cable changes in amplitude and speed in relation to a small change in the amplitude of the external force.

In this model $\delta_1$ and $\delta_2$ represent the viscous damping, $a_1$ the stiffness due to the cable, $a_2$ the stiffness due to the roadbed, $k$ the nonlinear coupling between the cable and the roadbed, $\varepsilon$ the amplitude of the external force on the cable, and $W_0$ the weight of the bridge, all normalized by the masses. In this model, the system parameters depend on both the mass of the cable ($m_1$) and the roadbed ($m_2$). As a result, one would expect $k_1$ and $a_1$ to be an order of magnitude larger than $k_2$ and $a_2$. In the simulations, the selected parameters are as shown in [11], where: $k_1$ = 10.0 [$N/mkg$], $a_1$ = 10.0 [$N/mkg$], $k_2$ = 1.0 [$N/mkg$], $a_2$ = 0.1 [$N/mkg$], $\delta_1$ = $\delta_2$ = 0.01 [$Ns/mkg$] and the initial conditions $z(0)$ = 5, $y(0)$ = $-5$, $\dot{y}(0)$ = (0) = 0 , for all three analyses studies.

In the first and second analyses there will be a direct comparison between the two, with the same frequency $\mu$ = 4.25 [$rad/s$], but with a difference in the $\varepsilon$ parameter. In the first analyse $\varepsilon$ = 0.3 [$N/kg$] and in the second analyse $\varepsilon$ = 0.4 [$N/kg$]. In the Figures 10((a), (b)), it is possible to see that the change in $\varepsilon$ leads the cable and roadbed to oscillate with larger amplitudes, however, remaining periodic as shown in the Figures 11((c),(d),(e),(f)), in which the last 500 [$s$] represent ellipses without any crossing of lines (periodic).

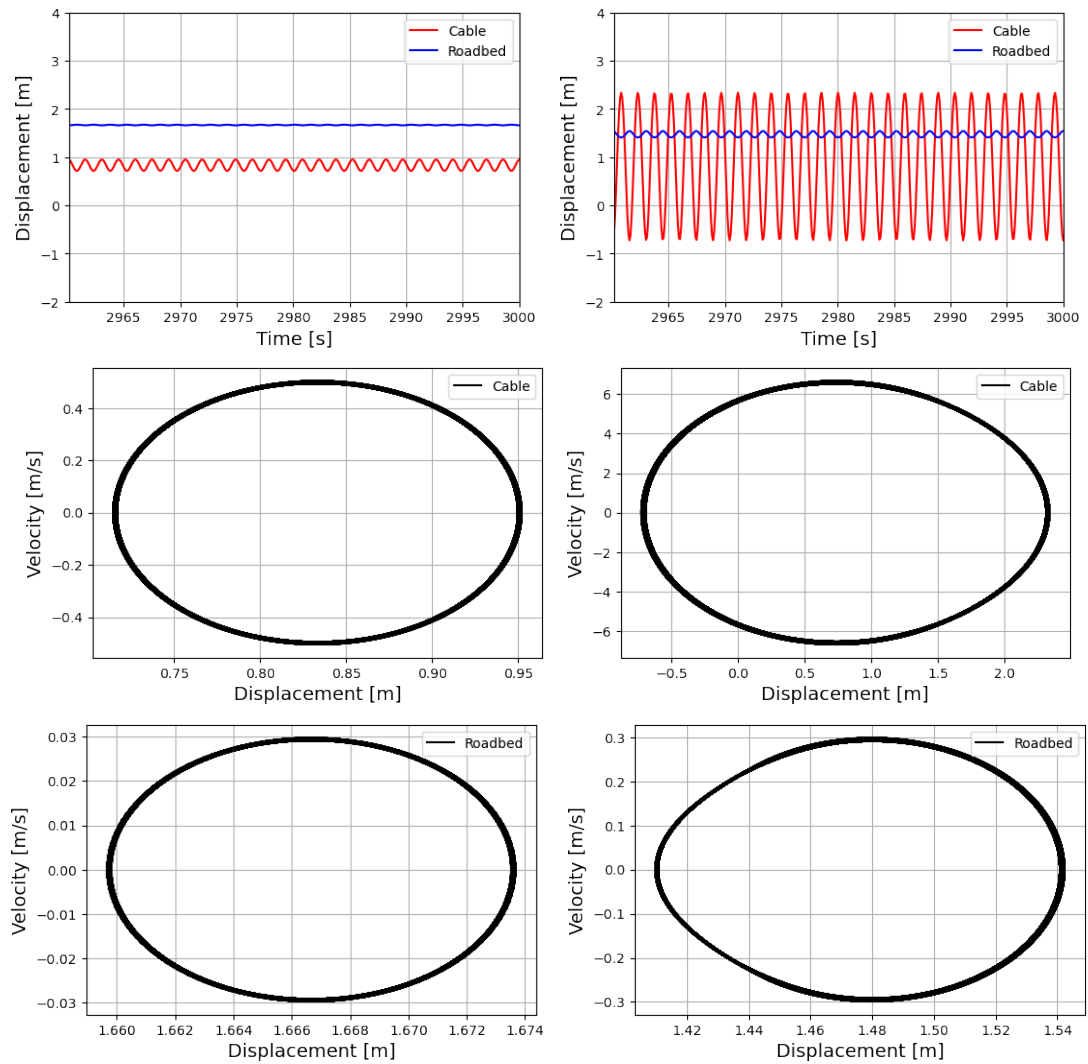

Figure 11. (a)Time domain response (λ=0.3), (b)Time domain response (λ=0.4), (c)Phase Portrait cable (λ=0.3), (d)Phase Portrait cable (λ=0.4), (e)Phase Portrait roadbed (λ=0.3), (f)Phase Portrait roadbed (λ=0.4).

In the analysis using FFT (Figure 12) and CWT (Figure 13) it is possible to observe a superharmonic appearing in the analyse where $\varepsilon$ = 0.4. In this, it is possible to observe the non-linear effect of galloping cables [9], where the cable moves with a relatively high amplitude and with a low frequency, around 1[$Hz$].

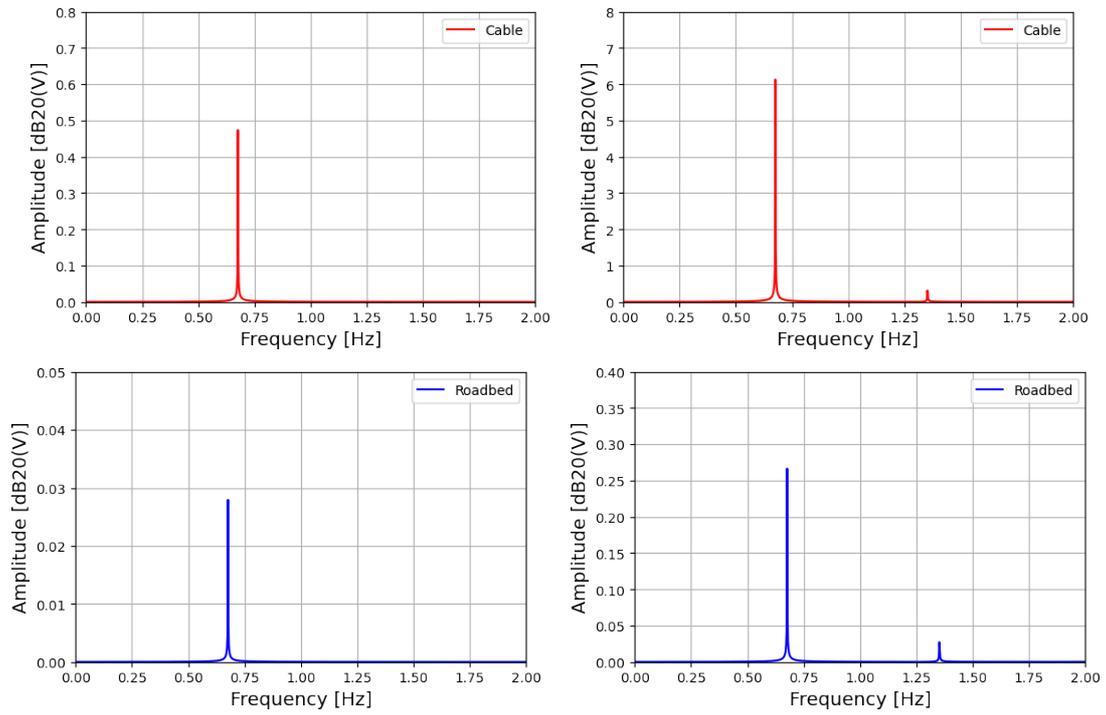

Figure 12. Frequency domain response (a)FFT cable (λ=0.3), (b)FFT cable (λ=0.4), (c)FFT roadbed (λ=0.3), (d)FFT roadbed (λ=0.4).

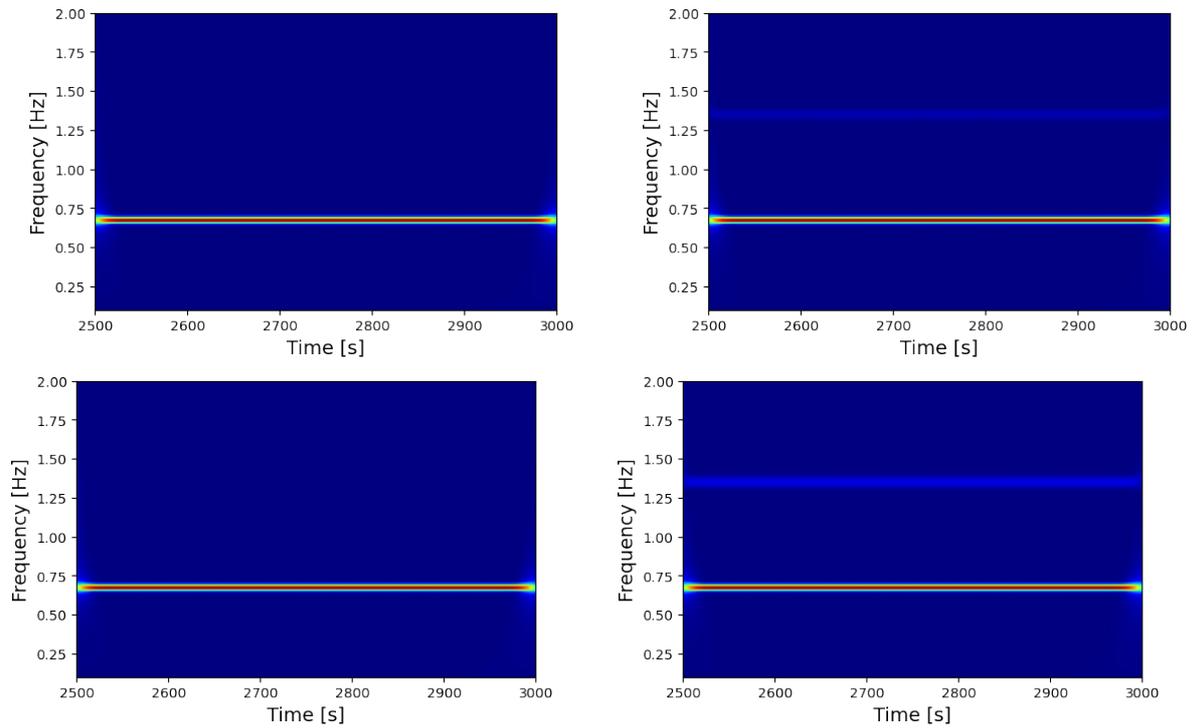

Figure 13. Frequency domain response (a) CWT Cable ($\lambda$=0.3) , (b) CWT Roadbed ($\lambda$=0.3), (c) CWT Cable ($\lambda$=0.4) (d) CWT Roadbed ($\lambda$=0.4).

In the third analyse, the effect that occurs when further increasing the parameters to $\mu = 4.5$ [$rad/s$] and $\varepsilon = 2.3$ will be shown. Here it is possible to observe the beating effect in Figure 14 (a), and large amplitudes for the roadbed with a dynamics with multiple periods, visualized in Figure 14 (b), plus in the phase spaces (Figures 14 ((c),(d)) ) it is possible to find solutions that intersect, typical of multi-periodic motions.

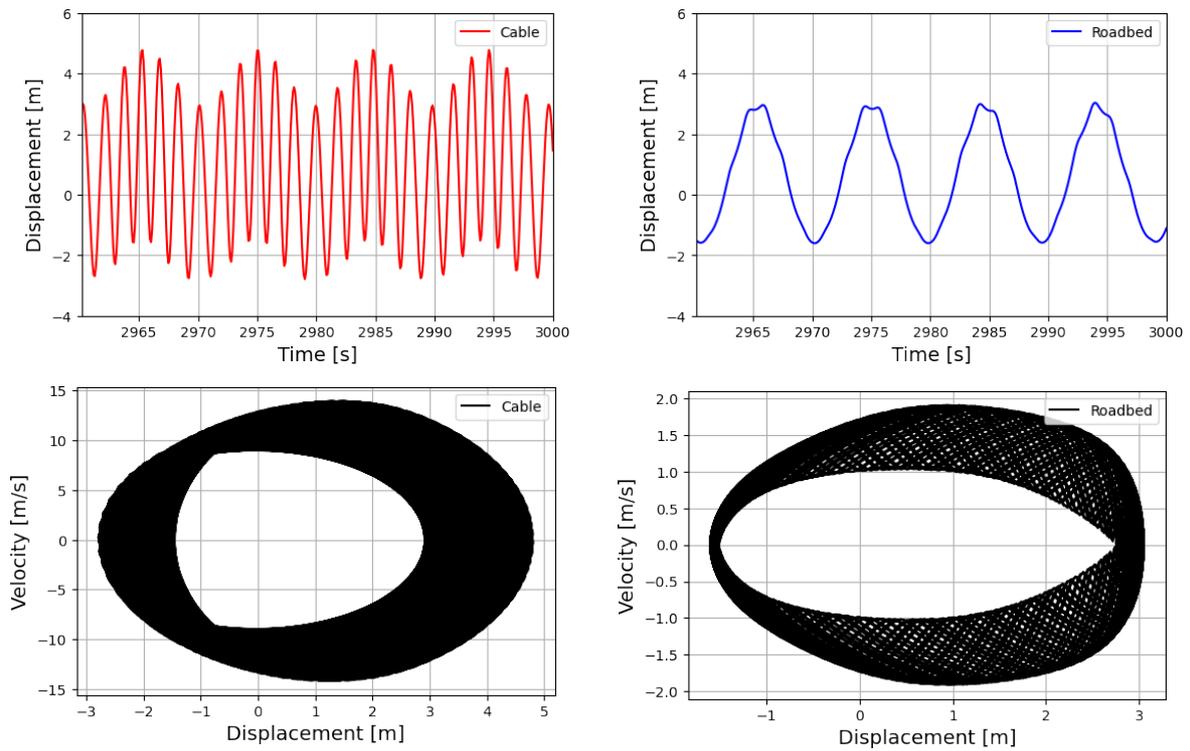

Figure 14. Time domain response (a)Time domain response cable (λ=2.3), (b)Time domain response roadbead (λ=2.3), (c)Phase portrait cable (λ=2.3), (d)Phase portrait roadbead (λ=2.3).

Time-frequency analysis, the presence of several harmonic frequencies is observed in the Figures 15 ((a),(b)), through FFT and in the Figures 15 ((c),(d)), through CWT.

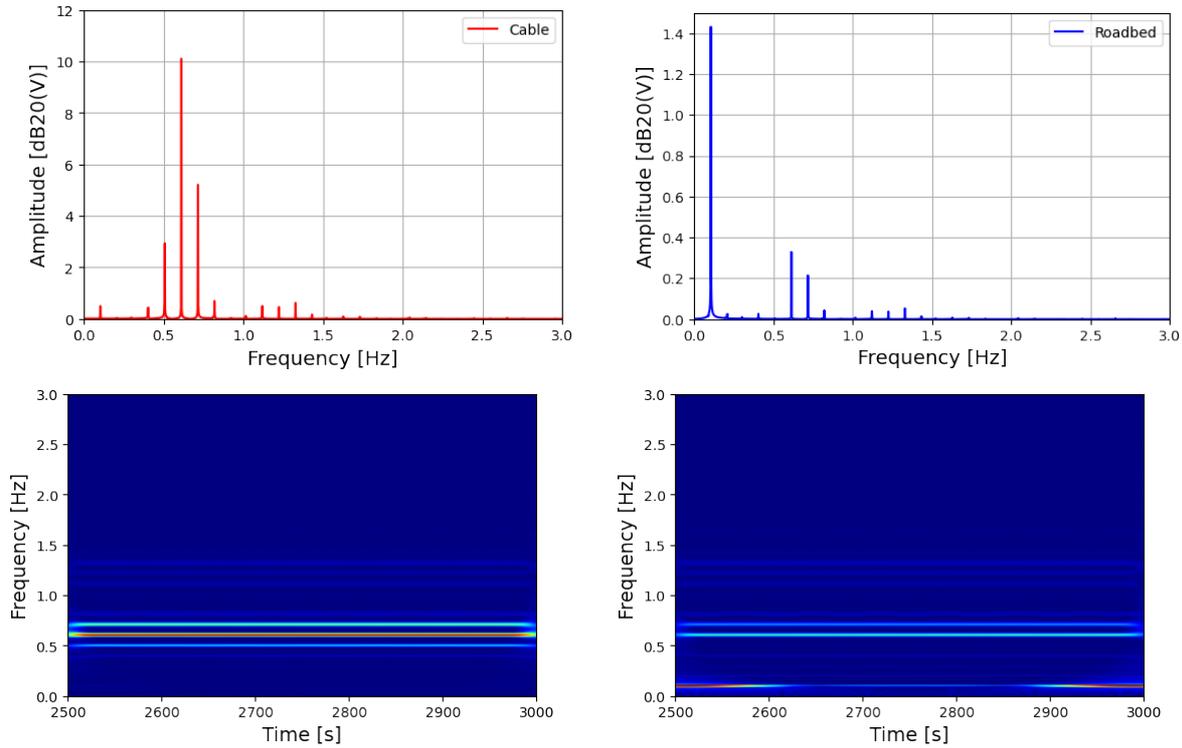

Figure 15. Frequency domain response (a) FFT cable ($\lambda$=2.3) (b) FFT roadbead ($\lambda$=2.3), (c) CWT cable ($\lambda$=2.3) (d) CWT roadbead ($\lambda$=2.3).

The harmonic frequencies observed in the frequency spectra come mainly from the mechanical nonlinearities of the system.

### 3.3. Case 3 - McKenna torsional oscillation.

This model has as main objective the idealization of vertical oscillations of a bridge deck cross section, besides taking into account a momentary loss of cable tension, which would lead to an instability in the torsional response.

By means of two coupled differential equations Eq.(9) and (10), one representing the vertical oscillations and the other, the torsional oscillations; a study of the difference caused by initial conditions will be presented, in its two analyses will be presented: the first where the initial vertical displacement will be 26 [$m$], and the second being 29 [$m$].

The viscous damping is denoted by $\delta$, spring constant $k$, mass $m$, bridge length $l$, amplitude of the external force $\lambda$, and frequency of the external force $\mu$. The parameters used for the simulation of this model were obtained from [12], and are therefore as follows: $\delta = 0.01$ [$Ns/m$], $k = 1000$ [$N/m$], $m = 2500$ [$kg$], $l = 6$ [$m$], $\lambda = 0.02$ [$N$], $\mu = 1.4$ [$rad/s$] e $g = 9.81$ [$m/s^2$]. Unlike the other the previous analyses, here the transient period will be also observed. Figure 16 shows the the time domain response.

For the analyse 1 where $y_0 = 26$ (Figure 16 (a)), the vertical oscillations drop until they are practically zero. Figure 16 (c) shows the torsional oscillations of the bridge, at around $3^o$ of maximum amplitude. Figure 16 (e) shows the transient behavior of the torsional oscillations and their steady state.

In analyse 2 where $y_0 = 29$, the same behavior of the vertical oscillations found in analyse 1 is observed, reaching almost zero. However, in the torsional oscillations, the vibration amplitudes suffered a significant change in relation to analyse 1, starting from around $3^o$, reaching around $35^o$ (Figure 16 (d)). Its Phase Portrait undergoes a clear change as well, not only in its steady state, but also in the transient behavior, observed in Figure 16 (f). Thus, a small change in the initial condition brought a big change in the system's response [30, 31, 32].

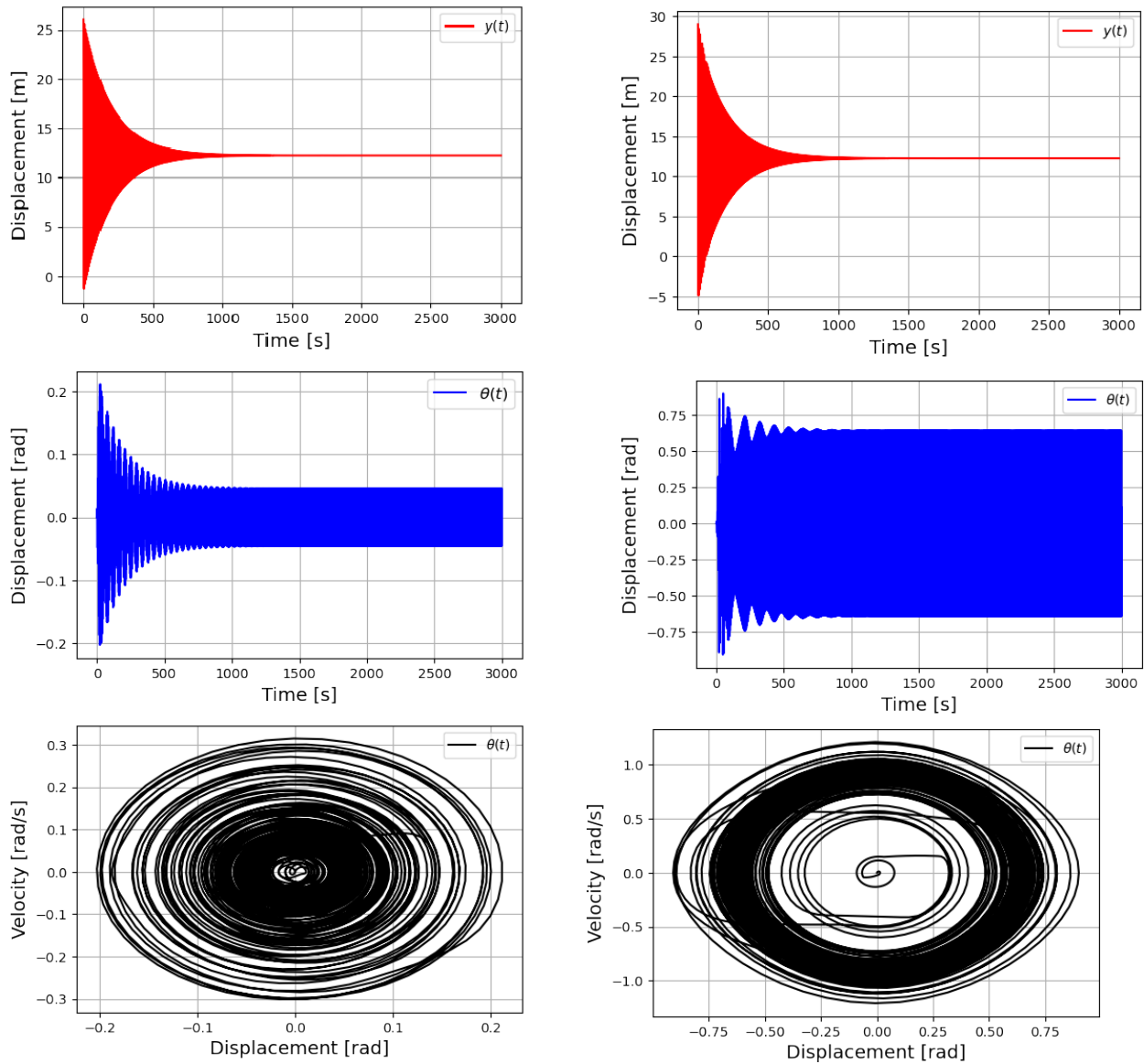

Figure 16. Time domain response (a) Time domain response ($y_0 = 26$) vertical, (b) Time domain response ($y_0 = 29$) vertical , (c) Time domain response ($y_0 = 26$) Torsional, (d) Time domain response ($y_0 = 29$) Torsional , (e)Phase portrait ($y_0 = 26$)Torsional , (f) Phase portrait($y_0 = 29$) Torsional.

For an study without the transient period, again the time interval between $2500 \geq t \geq 3000$ seconds will be taken, as presented in Figure 17. In the Figures 17 ((a), (c)) the analyse 1 ($y_0 = 26$) proved to be completely periodic with an amplitude around $3^o$. In analyse 2 ($y_0 = 29$), in the Figures 17 ((b), (d)), the same periodic behavior is found, but a high in vibration amplitudes is confirmed, reaching up to around $35^o$.

The results presented show that the vertical amplitudes become less intense with time, tending to zero, however, the torsional amplitudes continue oscillating with significant amplitudes. For a different initial condition, it is possible to observe how this change can lead to catastrophic cases of continuous torsional oscillations, characteristic of nonlinear systems.

In the case of the Tacoma Norrows Bridge failure report [12, 7], this almost instantaneous shift from vertical to torsional oscillations was observed, where the bridge remained with torsional oscillations for approximately 45 minutes until its collapse.

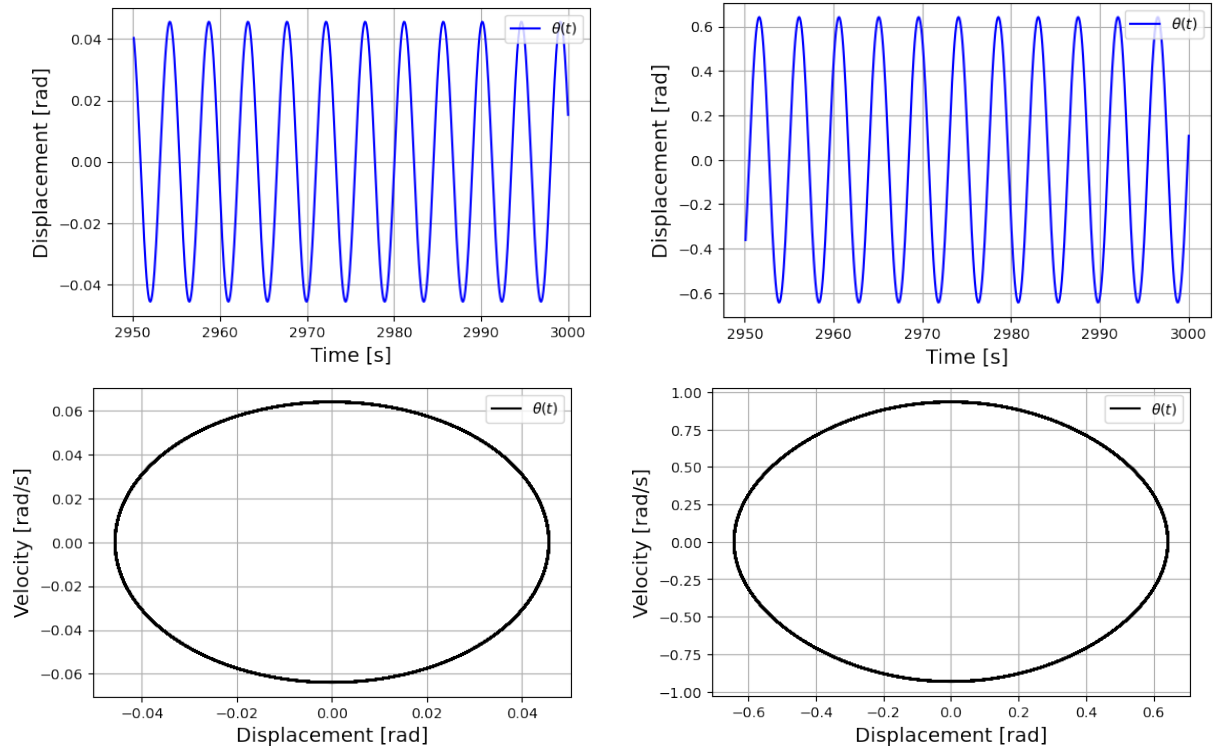

Figure 17. Time domain response (a) Time domain response ($y_0 = 26$), (b) Time domain response ($y_0 = 29$) , (c) Phase portrait ($y_0 = 26$), (d) Phase portrait($y_0 = 29$) .

The time-frequency analysis of the signal in the analyse 1 (Figure 18 ((a),(c))), shows stationary and linear behavior, but in the analyse 2 (Figure 18 ((b),(d))), a periodic response is obtained, but with a superharmonic [33, 34, 35, 36], a common effect presented in nonlinear systems [37, 38, 39].

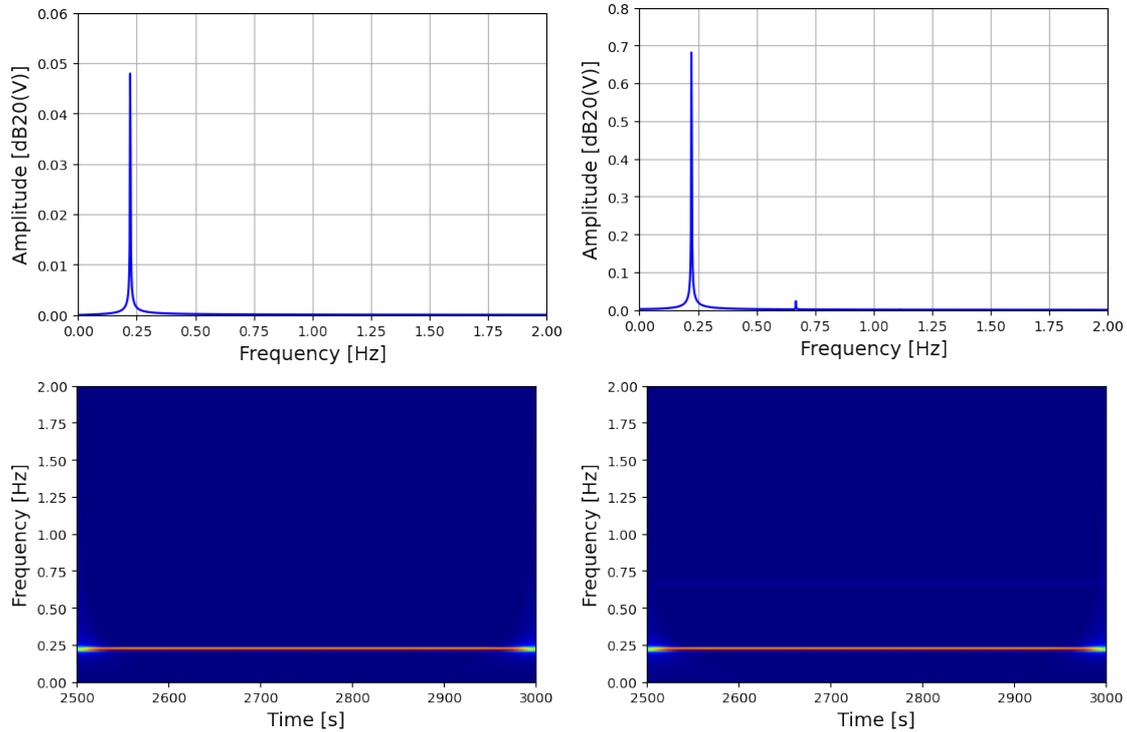

Figure 18. : Frequency domain response (a) FFT ($y_0$ = 26) (b) FFT ($y_0$ = 29), (c) CWT ($y_0$ = 26) (d) CWT ($y_0$ = 29).

## 4. Final remarks

The systems idealized by Lazer-McKenna to explain the dynamic behavior of suspension bridges, can characterize the nonlinearities found in the oscillations, section 3.1 Numerical simulations show that the analyzed system may or may not exhibit periodicity. For the analyse 1 where $\lambda$ = 7 [$N/kg$], a single frequency around 0.122[$Hz$] was found in accordance with linear theory, also characterized in the frequency domain through analysis with FFT, CWT and HHT, plus in Poincaré Map a single period was shown. In the analyse 2 where $\lambda$ = 15.5 [$N/kg$], nonlinear responses were observed in the time-frequency analysis are found in the system. In addition to the multiple frequencies present in the FFT, CWT and HHT, the Poincaré Map and Lyapunov exponents exhibits multiple periods, evidence that demonstrates chaotic behavior and system exhibits extremely rich dynamic behaviors.

For the second model, section 3.2, the external force parameters resulted in a significant change in the temporal response. Changing the $\varepsilon$ parameter caused an increase in amplitude in the temporal response, a more significant change for the main cable, which resembles the galloping cables phenomenon. With the change of $\varepsilon$ and $\mu$, the model showed a beat-like oscillation in the main cables, in addition the caracteriztion of multiple harmonics in the time-frequency analysis. Finally, the effectivenes of the theoretical results is showed through of numerical simulations.

In the last model, section 3.3, the changes from vertical to torsional oscillations were found through simulations, which would explain why the Tacoma bridge oscillations had the almost instantaneous transition from vertical to torsional oscillations, with these being the subsequent oscillations until its collapse.